\documentclass[epjCONF]{svjour}

\usepackage[english]{babel}
\usepackage{graphicx}
\usepackage[varg]{txfonts} 
\usepackage[latin1]{inputenc}

\graphicspath{{img/}}

\session-title{International Conference on New Frontiers in Physics (ICFP) 2012}

\begin{document}

\title{The Old New Frontier: Studying the CERN-SPS Energy Range with
  NA61/SHINE}
\author{Marek Szuba\inst{1}\fnmsep\thanks{\email{Marek.Szuba@kit.edu}} for the
  NA61/SHINE Collaboration\inst{2}}
\institute{Karlsruhe Institute of Technology, Germany \and CERN, Geneva,
  Switzerland}

\abstract{
With the Large Hadron Collider entering its third year of granting us insight
into the highest collision energies to date, one should nevertheless keep in
mind the unexplored physics potential of lower energies. A prime example here
is the NA61/SHINE experiment at the CERN Super Proton Synchrotron. Using its
large-acceptance hadronic spectrometer, SHINE aims to accomplish a number of
physics goals: measuring spectra of identified hadrons in hadron-nucleus
collisions to provide reference for accelerator neutrino experiments and
cosmic-ray observatories, investigating particle properties in the large
transverse-momentum range for hadron+hadron and hadron+nucleus collisions for
studying the nuclear modification factor at SPS energies, and measuring
hadronic observables in a particularly interesting region of the phase diagram
of strongly-interacting matter to study the onset of deconfinement and search
for the critical point of strongly-interacting matter with nucleus-nucleus
collisions. This contribution shall summarise results obtained so far by
NA61/SHINE, as well as present the current status and plans of its
experimental programme.
}

\maketitle

\section{\label{sec:intro}Introduction}

There is no doubt that studies of particle collisions at the highest available
energies can and do result in important scientific results; the string of
discoveries from the Tevatron at Fermilab and the recent observation at the
Large Hadron Collider at CERN of what is likely the Higgs boson attest to that.
Even so, one should keep in mind the unexplored physics potential of collisions
at lower energies. Data from experiments such as those at the Super Proton
Synchrotron (SPS) at CERN, operating well below the high-energy frontier,
continue to contribute to expanding our knowledge of matter, our world and the
universe. Examples of projects depending on such data include:

\subsection{\label{sec:infoNuOscillation}Measurement of Neutrino Oscillation}

Predicted by Pontecorvo 1957 and since then observed experimentally, neutrino
oscillation is a quantum-mechanical phenomenon which can make a neutrino
produced with a specific flavour be observed later as possessing different
flavour~\cite{Dore:2008dp}. It stems from differences in quantum-phase
propagation of different neutrino mass eigenstates and as such requires
neutrinos to have mass. One way to obtain accurate measurements of neutrino
oscillation is to perform it on neutrinos produced in a controlled environment
of a particle accelerator: a high-energy beam of protons is collided against
a graphite target to produce positive pions and kaons, which subsequently
decay into muons and muon neutrinos.

One experiment measuring oscillation of beam neutrinos is T2K in Japan,
directing particles from the J-PARC facility in Tokai towards the
Super-Kamiokande detector 295~km away in Kamioka~\cite{Itow:2001ee}. Physics
goals of T2K are to obtain one of the first measurements of the $\theta_{13}$
mixing angle of the Pontecorvo-Maki-Nakagawa-Sakata (PMNS) neutrino-mixing
matrix, to improve precision of measurements of the $\theta_{23}$ mixing angle
and the mass difference $\Delta m^{2}_{23}$, and in the future to search for
$\nu$ CP violation. Other experiments of this sort include MINOS and NO$\nu$A
(NuMI beam at Fermilab), LBNE (upcoming new $\nu$ beam at Fermilab), and OPERA
and ICARUS (CNGS beam at
CERN)~\cite{Ables:1995wq}\cite{Ayres:2004js}\cite{Sanchez:2009zzf}\cite{Guler:2000bd}\cite{Arneodo:2001tx}.

Unfortunately even with well-defined primary beams, estimating the resulting
neutrino flux is not a trivial matter. In order to achieve adequate beam
intensity the graphite target must be long, leading to secondary interactions.
Moreover, a non-negligible contribution to the flux is made by interactions
with support structures such as the target holder
and cooling systems. In short, good knowledge of properties of the resulting
neutrino beam requires good knowledge of hadron production in the target ---
which in turn may be much easier to study using an identical target at a
fixed-target facility oriented towards hadron spectroscopy than having to set
up necessary detectors \textit{in situ}. Given the energies of proton beams
typically used for production of neutrino beams, the CERN SPS complex (indeed,
in case of the CNGS beam the proton accelerator in question \emph{is} the SPS)
is a perfect candidate for this task.

\subsection{\label{sec:infoEAS}Hadron Production in Extensive Air Showers}

Another domain which can benefit from hadroproduction measurements in the
energy range of the SPS are studies of extensive air showers (EAS) produced in
Earth's atmosphere by cosmic rays. Detecting and measuring EAS is the standard
technique for studying ultra high-energy cosmic rays, as such particles reach
our planet so infrequently and from so many different directions that it is virtually
impossible to gather a statistically significant sample through direct
measurements (\textit{i.e.} using satellite- or balloon-based detectors to
observe cosmic-ray particles themselves); using EAS detectors makes it both
technically and financially easier to cover a much wider area. Examples of
modern EAS experiments include KASCADE-Grande in Germany, Pierre Auger
Observatory in Argentina and the Telescope Array in the United
States~\cite{Badea:2004un}\cite{Abraham:2004dt}\cite{Martens:2007ef}.

Improved observation rates of EAS-based cosmic-ray experiments come at
a cost of having to accurately reconstruct properties of the original particle
from what has undergone multiple levels of interactions with the atmosphere
before reaching the detectors. In particular, evolution of the \emph{hadronic}
component of showers remains relatively poorly understood. With primary
energies of ultra high-energy cosmic rays remaining largely out of range of
man-made accelerators, one is required to employ simulations to provide
reference for such evolution --- which in turn requires careful tuning of
models used for this purpose.

This is where measuring hadron spectra at the SPS comes in. Although its
energy range falls several orders of magnitude short of the energy of cosmic
rays observed by EAS detectors ($E_0 = 10^{15} - 10^{20}~\mathrm{eV}$, it
provides an excellent match to the energy of the last generation of hadronic
interactions in the shower. As many experimental observables (\textit{e.g.} the
number of muons detected by surface detectors) are directly tied to this stage
of shower evolution, studying it at the SPS is expected to yield significantly
improved predictions.

\subsection{\label{sec:infoHighPT}High-$p_{T}$ Hadrons in p+p and p+A Colliisions}

Proton-proton and proton-nucleus collisions constitute important reference
systems for a wide range of different studies (for instance spectra, scaling
with the number of wounded nucleons or binary collisions, nuclear modification
factor and Cronin effect, among others) in nucleus-nucleus reactions. However,
in the past data sets of this sort at SPS energies and below were relatively
small. This problem becomes particularly pressing in case of high-$p_{T}$
hadrons, originating from hard scatterings and thus useful for probing the
perturbative-quantum chromodynamics (QCD) regime of strong interactions ---
they are produced so rarely that most such elementary collisions contain none
at all, further reducing effective sample size. Fortunately, development of
sophisticated detectors and readout components capable of coping with high
event rates of high-luminosity machines such as the LHC has made it possible to
also improve event rates of lower-energy detectors, allowing at last collection
of large-statistics p+p and p+A data sets. Analysis of this data is expected to
greatly improve precision of many existing results as well as allow new,
hitherto-infeasible studies.

\subsection{\label{sec:infoCPoD}Critical Point and the Onset of Deconfinement of QCD Matter}

Theoretical predictions based on QCD tell us that above certain energy density,
quarks and gluons previously confined to hadrons can undergo a phase transition
into the quark-gluon plasma --- a state in which they can be considered free.
The energy range of the SPS is very important for studying this transition
because at these energies one can produce collisions both right below and right
above the energy threshold for deconfinement~\cite{Alt:2007aa}. Moreover, it is
also in this energy range that we now expect the critical point of the QCD
phase diagram~\cite{Fodor:2004nz}.

A number of experimental observables were proposed to be sensitive to
the onset of deconfinement; these include changes in energy dependence of the
pion yield per wounded nucleon (``the kink''), of the $\frac{\left< K^+
\right>}{\left< \pi^+ \right>}$ ratio (``the horn'') and of the mean transverse
mass (``the step'')~\cite{Gazdzicki:1998vd}. On the other hand, the emergence
of the critical point is predicted to be observable in event-by-event
fluctuations of \textit{e.g.} multiplicity and average transverse
momentum~\cite{Stephanov:1999zu}. Indications in favour of this hypothesis has
been observed both at the SPS and
elsewhere~\cite{Alt:2007aa}\cite{Grebieszkow:2009jr}\cite{Kumar:2011us}.

\section{\label{sec:shineExperiment}The NA61/SHINE Experiment}

\subsection{\label{sec:shineOverview}Overview and Physics Goals}

NA61/SHINE (\textbf{S}PS \textbf{H}eavy \textbf{I}on and \textbf{N}utrino
\textbf{E}xperiment) is a fixed-target experiment located in the North Area of
the CERN SPS, using a large-acceptance hadronic spectrometer to study a wide
range of phenomena in a number of different hadron-hadron, hadron-nucleus and
nucleus-nucleus reactions~\cite{Antoniou:2006mh}. It is the successor of the NA49
experiment, which took data in the years 1994--2002, and reuses most of its
predecessor's hardware and software~\cite{Afanasev:1999iu}. Its large
acceptance (around 50~\% for $p_{T} \le$ 2.5~GeV/c), high momentum
resolution ($\sigma(p)/p^{2} \approx 10^{-4}~(\mathrm{GeV/c})^{-1}$) and tracking
efficiency (over 95~\%), and excellent particle-identification capabilities
($\sigma(\frac{\mathrm{d}E}{\mathrm{d}x}) / \frac{\mathrm{d}E}{\mathrm{d}x}
\approx 4~\%, \sigma(t_{ToF}) \approx 100~\mathrm{ps}$) make it an excellent
tool for investigating hadron spectra.

\begin{figure}
  \begin{center}
    \includegraphics[width=0.9\columnwidth]{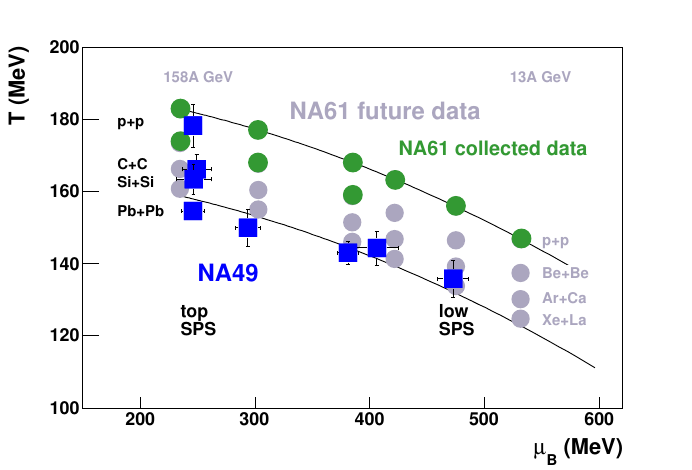}
  \end{center}
  \caption{\label{fig:2dScan}Estimated~\cite{Becattini:2005xt} (squares) and
  extrapolated (circles) chemical freeze-out points corresponding to NA49 and
  NA61/SHINE for both collected and future data.}
\end{figure}

NA61/SHINE has the following physics goals:
\begin{itemize}
  \item search for the critical point on the phase diagram of
    strongly-interacting (QCD) matter and study in detail the onset of
    deconfinement by performing a two-dimensional scan of the phase diagram
    (see Figure~\ref{fig:2dScan}),
  \item investigate production of high-$p_{T}$ hadrons in p+p and p+Pb
    collisions,
  \item provide reference measurements for the T2K neutrino experiment by
    measuring hadron production in collisions of a proton beam at 31~GeV
    with a carbon platelet and a replica of the T2K target,
  \item provide reference measurements for current and future Fermilab
    neutrino experiments by measuring hadron production in collisions of a
    proton beam at 60, 90 and 120~GeV with a carbon platelet and a replica of the
    NuMI target,
  \item perform reference measurements of hadron production in p+p, p+C and $\pi$+C
    interactions, for cosmic-ray extended air-shower experiments: KASCADE,
    KASCADE-Grande and Pierre Auger Observatory.
\end{itemize}

\subsection{\label{sec:shineDetector}Detector Set-up}

\begin{figure}
  \begin{center}
    \includegraphics[width=0.9\columnwidth]{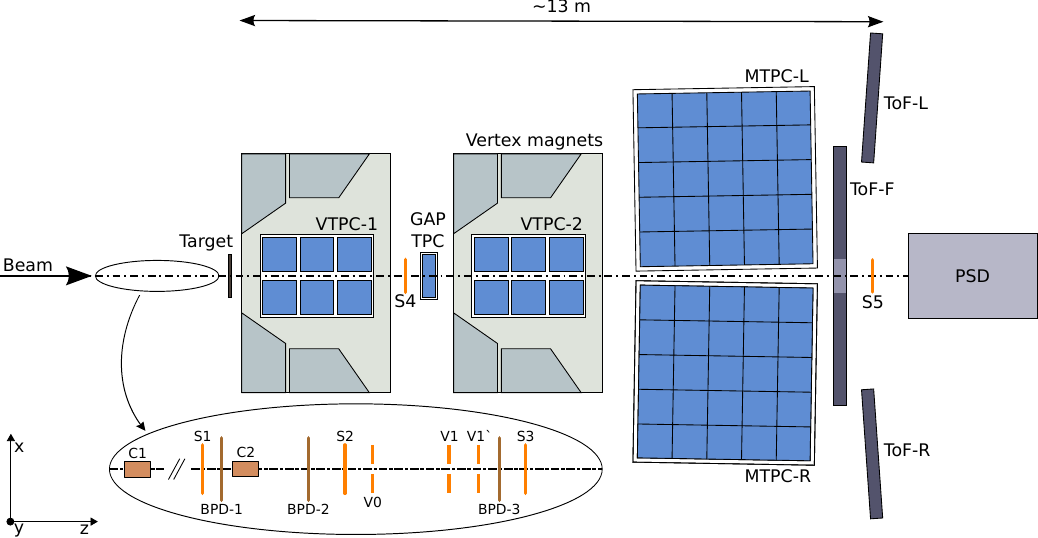}
  \end{center}
  \caption{\label{fig:shineDetector}Layout of the NA61/SHINE experimental
  set-up (top view, not to scale). For details see~\cite{Afanasev:1999iu}.}
\end{figure}

Figure~\ref{fig:shineDetector} shows a diagram of the NA61/SHINE apparatus. The
primary, tracking detectors are four large-volume Time Projection
Chambers (TPCs) inherited from NA49. Two of them (Vertex TPCs) are located inside
superconducting magnets along the beam line, the other two (Main TPCs) are placed
symmetrically to the beam line further downstream.  Another, smaller chamber
(Gap TPC) is placed between the VTPCs. Together, these chambers measure
trajectories and momentum of particles as well as allow their
identification through ionisation energy loss.

Just downstream of the MTPCs there exist three Time-of-Flight (ToF) walls, the side
two inherited from NA49 and the central one added by NA61 in 2007. Measurements
from these detectors complement particle-identification information in the
momentum range where $\frac{\mathrm{d}E}{\mathrm{d}x}$ alone is ambiguous.

The most downstream detector of the apparatus is the Projectile Spectator
Detector (PSD), installed in 2011. The purpose of this calorimeter is to
provide precise, high-granularity event-by-event measurements of the energy of
non-interacting fragments of the projectile nuclei, making it possible to
determine centrality of the collision and the orientation of the reaction
plane.

Finally, a number of additional detectors installed in the beam line both
upstream and downstream of the target monitor beam properties and provide
trigger information.

It is also worth mentioning at this point that a sophisticated beam line
set-up was designed and put in place by the CERN accelerators-and-beams
department in collaboration with NA61 for the purpose of providing beryllium
beams for SHINE. At the moment the only primary beams the SPS can provide
are protons and lead ions, with argon and xenon to be made possible during the
long shutdown of the accelerator complex in 2013-2014. As adding new species of
primary beams is an immensely complicated task, beryllium beams for SHINE were
instead produced in a fragmentation-ion beam line. Primary lead ions from
the SPS were collided with a beryllium block far upstream in the beam line and
the resulting fragments of the projectile were guided through a sequence of
magnets, collimators, degraders and beam detectors in order to filter out
beryllium ions.
This configuration was proven to produce highly pure, stable beams of $^7$Be
over a wide range of Pb-beam energies.

\subsection{\label{sec:shineDataTaking}Data Taking}

NA61/SHINE has been taking data since the year 2007 and so far has acquired
over 150 million events in a variety of data sets. We focused primarily on data
for cosmic-ray and neutrino physics in the first three years of running, and
progressed to high-$p_{T}$ data and the exploration of the phase diagram in the
following years. The presently-approved physics programme is foreseen to be
concluded in the year 2015.

\section{\label{sec:results}Results So Far}

\begin{figure}
  \begin{center}
    \includegraphics[width=0.85\columnwidth]{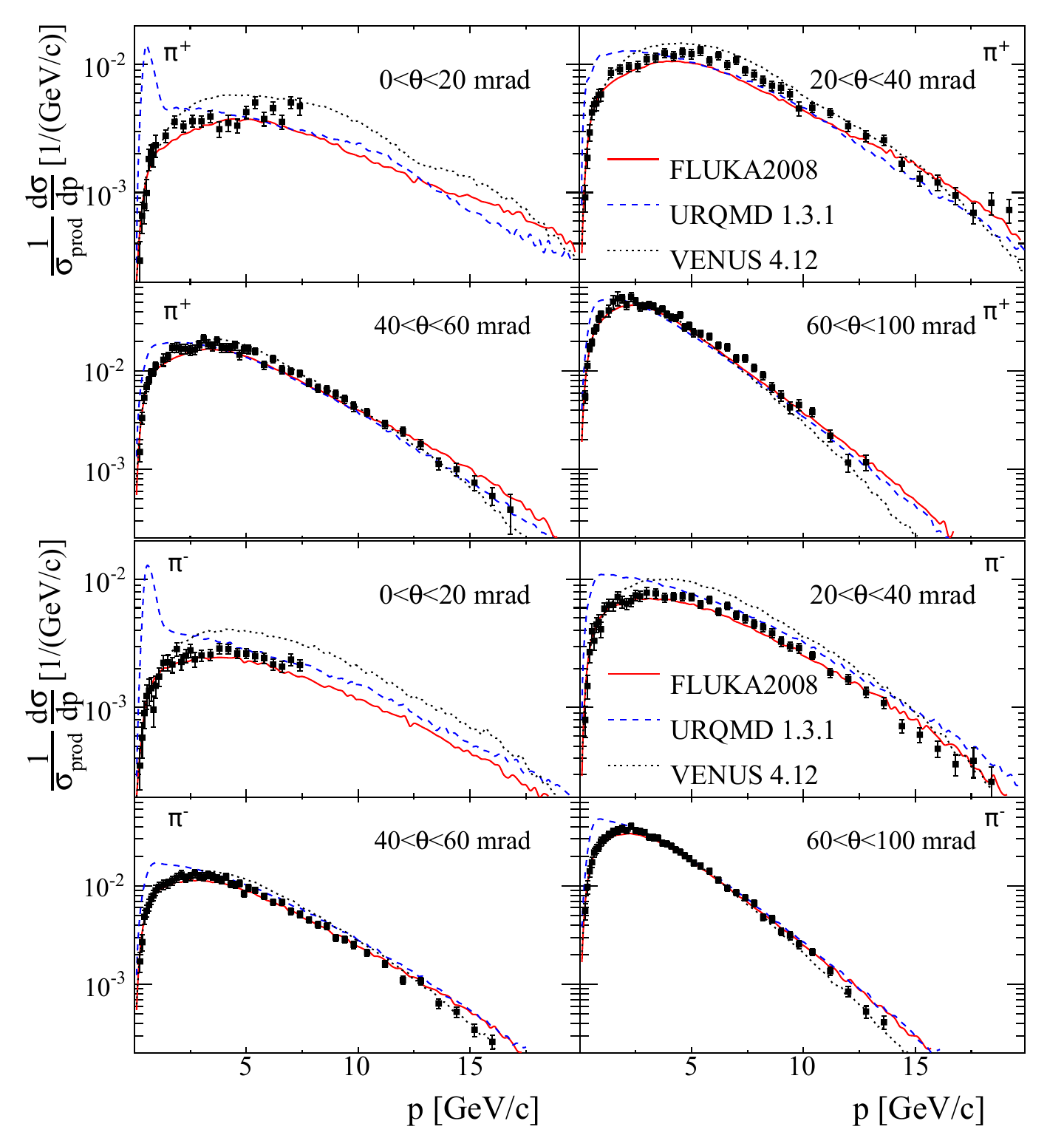}
  \end{center}
  \caption{\label{fig:p2C31pi}Inclusive spectra of positive (top) and negative
  (bottom) pions, in several polar-angle bins, measured by NA61/SHINE in p+C
  collisions at 31~GeV (points), compared to predictions (lines) from
  FLUKA~2008, UrQMD~1.3.1 and VENUS~4.12~\cite{Abgrall:2011ae}.}
\end{figure}

\begin{figure}
  \begin{center}
    \includegraphics[width=0.45\columnwidth]{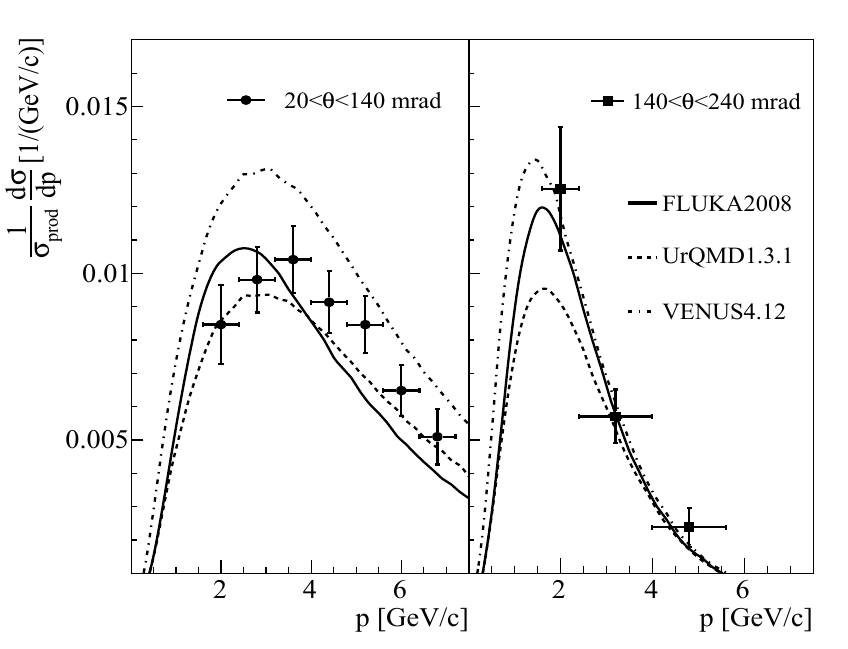}
    \includegraphics[width=0.45\columnwidth]{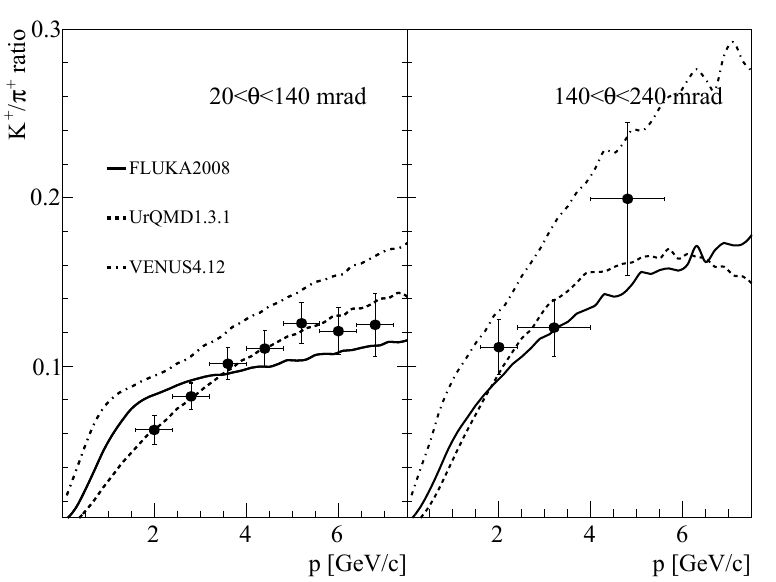}
  \end{center}
  \caption{\label{fig:p2C31K}\textbf{Left}: Inclusive spectra of
  positively-charged kaons measured by NA61/SHINE in p+C collisions at 31~GeV
  (points), compared to predictions by various models (lines).
  \textbf{Right}: Comparison of the $K^+/\pi^+$ ratio, using the above data
  along with results shown in Fig.~\ref{fig:p2C31pi}, again compared to
  models~\cite{Abgrall:2011ts}.}
\end{figure}

The first $\pi^\pm$ (see Figure~\ref{fig:p2C31pi}) and $K^+$
(Figure~\ref{fig:p2C31K}) spectra from p+C interactions at 31~GeV, based
on data acquired in 2007, are
published~\cite{Abgrall:2011ae}\cite{Abgrall:2011ts}. These results have
already been used by the T2K experiment as reference for calculations of
neutrino flux in the J-PARC beam, leading to their first measurement of the
$\theta_{13}$ $\nu$ mixing angle~\cite{Abe:2011sj}. Moreover, comparison of
pion spectra with models have made it possible to introduce improvements to
UrQMD as well as to the implementation of FRITIOF in the Geant4
toolkit~\cite{Uzhinsky:2011ir}\cite{Uzhinsky:2011qb}. Results from data taken
in 2009, offering a factor-of-ten increase in statistics as well improved
particle identification through extension of the forward Time-of-Flight wall,
are being prepared for release.

\begin{figure}
  \begin{center}
    \includegraphics[width=0.32\columnwidth]{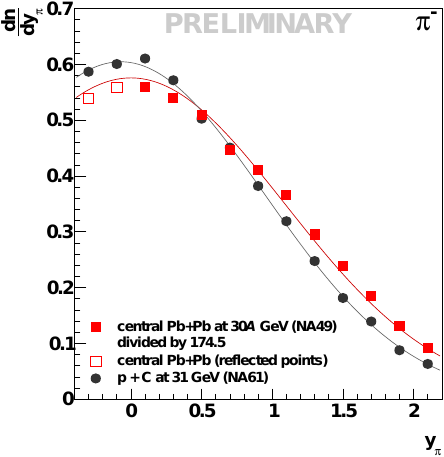}
    \includegraphics[width=0.32\columnwidth]{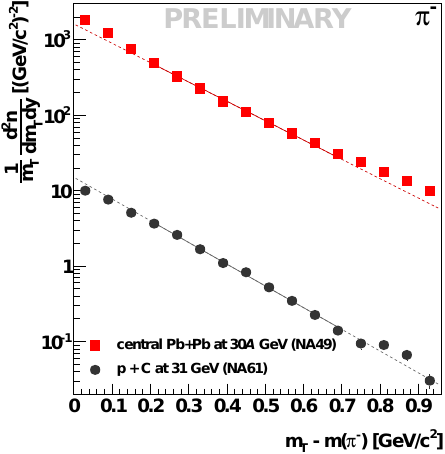}
    \includegraphics[width=0.32\columnwidth]{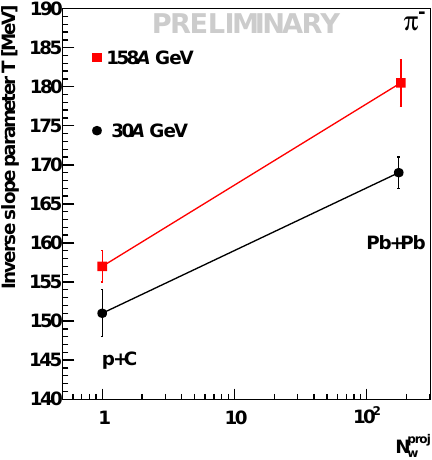}
  \end{center}
  \caption{\label{fig:piMinusExtraSpectra}\textbf{Left}: Rapidity spectra of
  $\pi^-$ in p+C collisions at 31~GeV and (scaled) central Pb+Pb collisions at
  30$A$~GeV, for $p_{T} < 1~GeV/c$. Corresponding gaussian fits are also shown
  as lines. \textbf{Middle}: Transverse-mass spectra of $\pi^-$ in the same
  reactions, for $0 < y < 0.2$. Dashed lines illustrate exponential fits to the
  data. \textbf{Right}: Inverse-slope parameters extracted from $m_{T}$ spectra
  for four different reactions measured by NA61 and NA49, with system size
  expressed as the number of wounded nucleons in the projectile.}
\end{figure}

Additional, preliminary results from p+C interactions at 31~GeV have been
compared to measurements of NA49 in Pb+Pb collisions~\cite{Grebieszkow:2012vj}.
Transverse-mass and
forward rapidity spectra of negative pions were extracted and compared to NA49
results from central Pb+Pb collisions at 30$A$~GeV (see
Figure~\ref{fig:piMinusExtraSpectra}). We have also begun investigating yields
of $\Lambda$, $K^0_S$ and $\Delta^{++}$ particles in these interactions.

\begin{figure}
  \begin{center}
    \includegraphics[width=0.8\columnwidth]{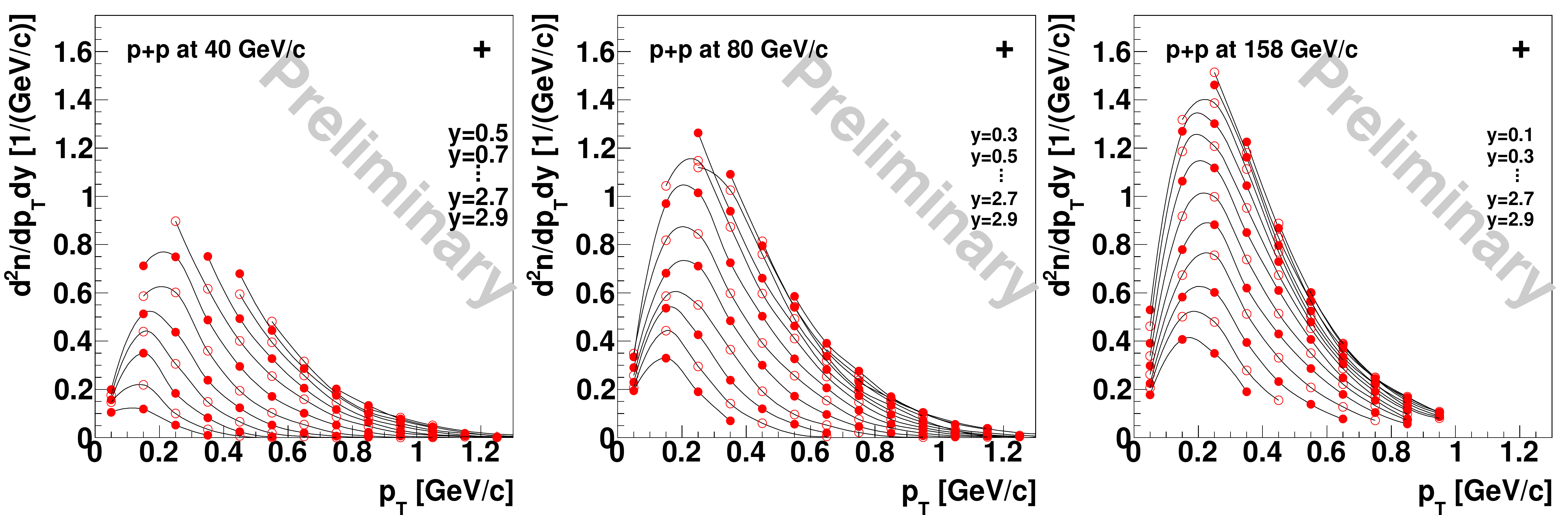}
    \includegraphics[width=0.8\columnwidth]{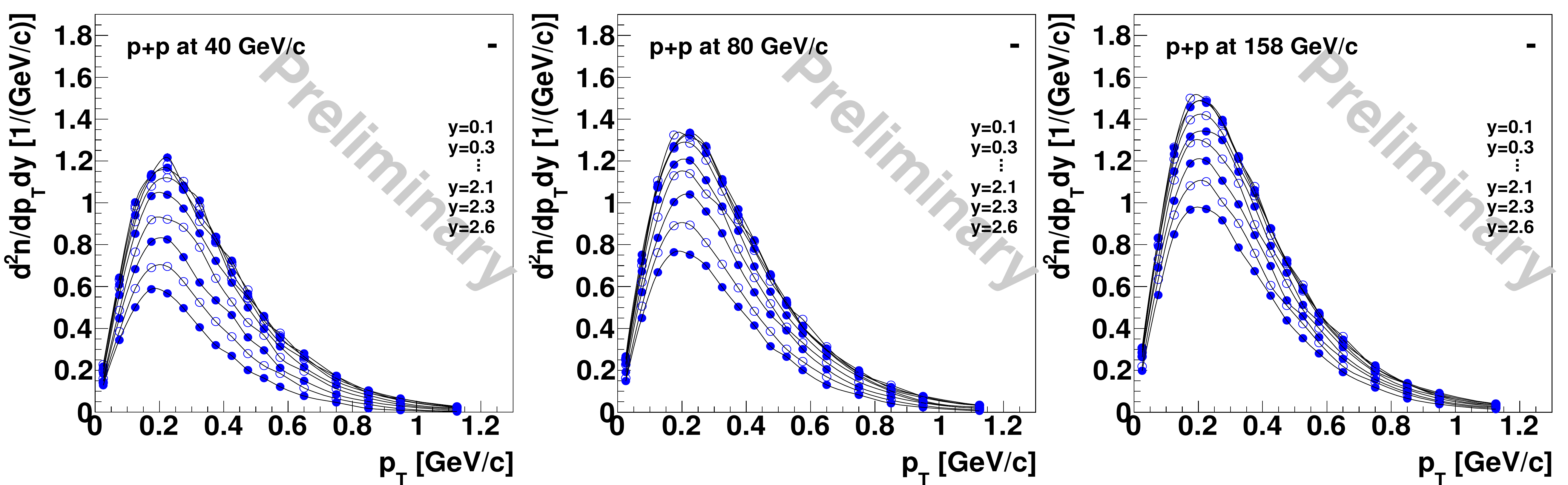}
    \includegraphics[width=0.8\columnwidth]{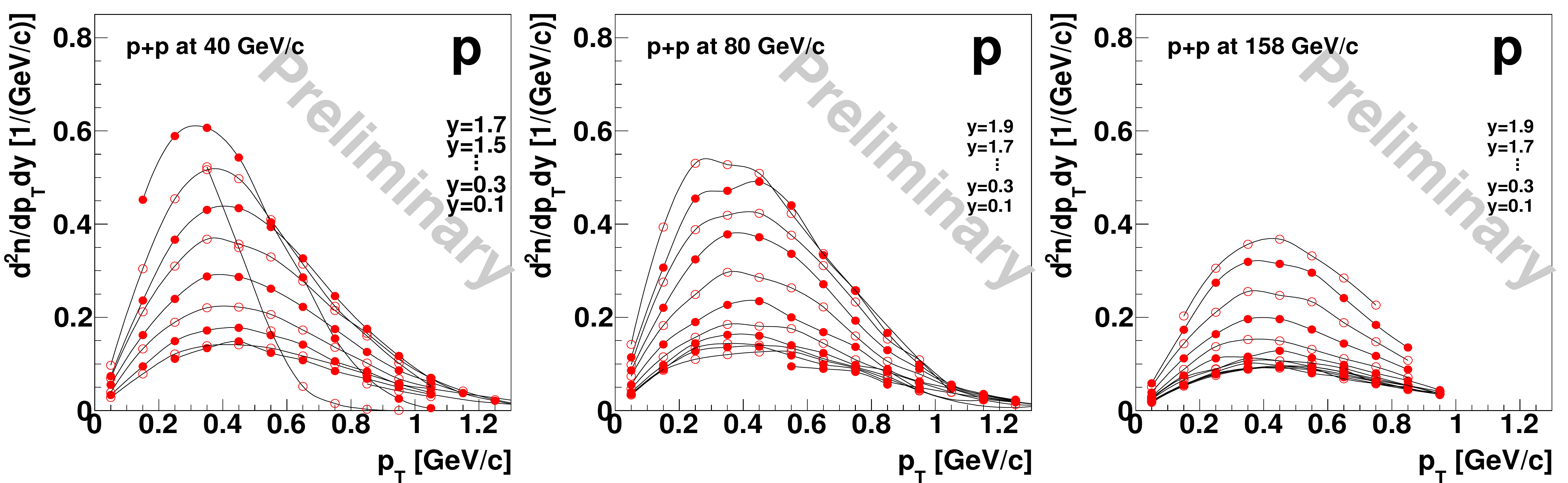}
    \includegraphics[width=0.8\columnwidth]{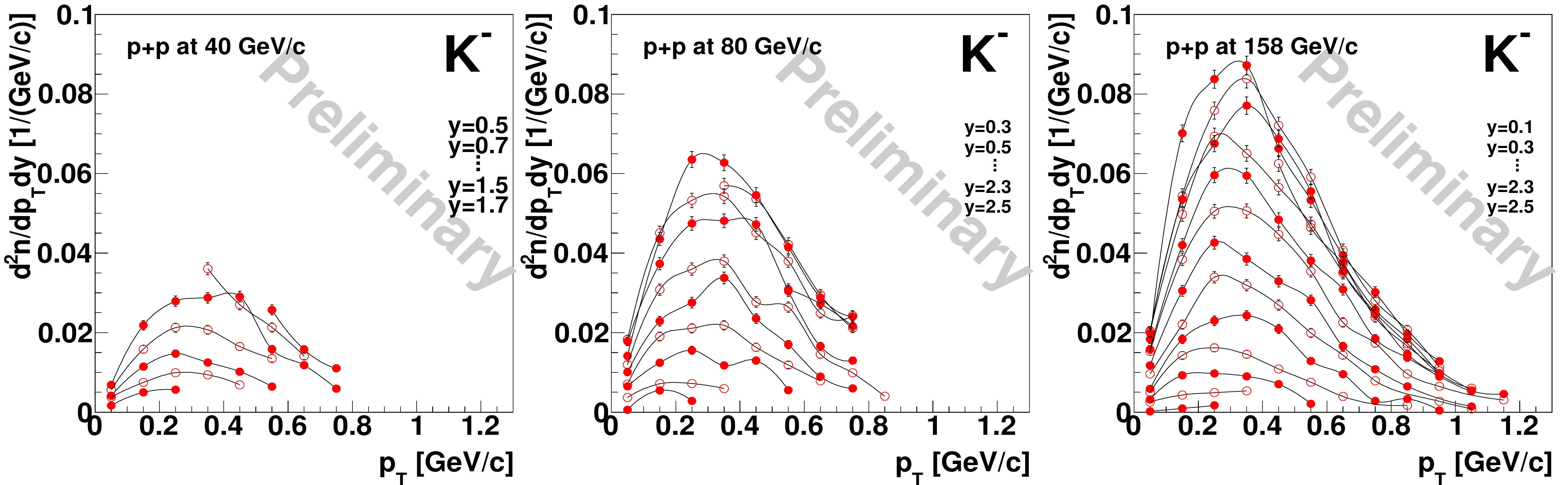}
  \end{center}
  \caption{\label{fig:ppSpectra}Double-differential spectra of $\pi^+$,
  $\pi^-$, $p$ and $K^+$ (from top to bottom, respectively) in
  inelastic p+p interactions at 40, 80 and 158~GeV. Only statistical
  uncertainties are shown; average systematic error is approximately
  8~\%~\cite{ppSpectraQM12}.}
\end{figure}

Last but not least, we have recently presented the first preliminary $\pi^\pm$,
$K^-$ and $p$ spectra from p+p interactions at 40, 80 and 158~GeV ---
reference data sets for the phase-diagram scan~\cite{ppSpectraQM12}; an
overview of these results can be found in Figure~\ref{fig:ppSpectra}.
Analogous measurements from p+p events at 13, 20 and 31~GeV are to follow suit.

\section{\label{sec:summary}Summary}

NA61/SHINE results have already provided important input for the calculation of
the neutrino flux in the T2K beam which was used for the measurement of
$\theta_{13}$. Moreover, models employed for simulating cosmic-ray showers were
significantly constrained. The scan of the phase diagram of strongly
interacting matter has begun and first results from p+p collisions were
presented.

\bibliographystyle{epj}
\bibliography{shine}

\end{document}